\documentstyle[prl, multicol, aps, epsf]{revtex}

\begin{document}

\draft

\title{\large \bf The Energy-dependent Checkerboard Patterns in Cuprate Superconductors}

\author{Degang Zhang and C. S. Ting}

\address{Texas Center for Superconductivity and Department of Physics,
University of Houston, Houston, TX 77204, USA}

\maketitle

\begin{abstract}

Motivated by the recent scanning tunneling microscopy (STM)
experiments [J. E. Hoffman {\it et al.}, Science {\bf 297}, 1148
(2002); K. McElroy {\it et al.}, Nature (to be published)], we
investigate the real space local density of states (LDOS) induced
by weak disorder in a d-wave superconductor. We first present the
energy dependent LDOS images around a single weak defect at
several energies, and then point out that the experimentally
observed checkerboard pattern in the LDOS could be understood as a
result of quasiparticle interferences by randomly distributed defects.
It is also shown that the checkerboard pattern oriented along $45^0$ to
the Cu-O bonds at low energies would transform to that oriented
parallel to the Cu-O bonds at higher energies. This result is
consistent with the experiments.

\end{abstract}


\begin{multicols}{2}

Recently the energy-dependent modulations of the local density of
states (LDOS) in high temperature superconductors has attracted a
lot of experimental and theoretical attentions[1-6]. By employing
the high-resolution Fourier-transform scanning tunneling
microscopy (STM), Hoffman {\it et al.} investigated the zero-field
charge modulations in Bi$_2$Sr$_2$CaCu$_2$O$_{8+\delta}$ [1]
deduced from the checkerboard patterns in the real space LDOS. The
charge modulation vectors near the origin of the momentum space
were determined. They found that the period of the modulations
depends on the energy and doping for energy  below the maximum
superconducting gap. With increasing energy (doping fixed) or
doping (energy fixed), the LDOS modulation wave vectors oriented
parallel to the Cu-O bonds become shorter while those along $45^0$
to the Cu-O bonds become longer. They also observed that when
energy increases, the charge modulation along $45^0$ to the Cu-O
bonds changes to that along the Cu-O bonds. Subsequently, McElroy
{\it et al.} extended the previous measurements [1] to the second
Brillouin zone and discovered the characteristic octet of
quasiparticle states [2, 4].

A number of theoretical studies have devoted to the explanation of
the STM experiments [3-6]. It has been proposed that the
experimental phenomenon is due to the result of quasiparticle
interference induced by disorder [1, 2]. Following this idea, Wang
and Lee calculated the Fourier component of the LDOS produced by
an single impurity with a moderate strength on-site potential [3].
They obtained the LDOS images in momentum space and compared them
with those in Ref. [1], but didn't examine the relations among the
modulation wave vectors, dopings and the bias voltages. The
present authors, on the other hand, investigated the effect of
quasiparticle scatterings from a weak and extended impurity or
defect [4]. The Fourier transform images of the LDOS and the
relations among modulation wave vectors at different dopings and
the bias voltages obtained by us [4] in the first Brillouin zone
are consistent with the experimental observations [1, 2]. There
exist also other works trying to understand the STM experiments
[5, 6]. However, almost all the previous studies [3,4,5] were
restricted to the discussion of the Fourier transform of the LDOS
due to a single defect in the first Brillouin zone. So far there
have existed no studies of the real space LDOS images at different
energies, and the origin of the checkerboard patterns observed
in the experiments. Thus it is necessary to do these calculations,
and to compare the obtained results directly with the LDOS images
in the STM experiments [1, 2].

In this paper, we base on the approach of our previous study [4]
and examine the effect due to quasiparticle scattering from weak
defects or impurities with both hopping and pairing modifications
on the LDOS. The reason we choose weak defects as scatters is
simply to eliminate the contribution from the resonant states.
Using the T-matrix approach, We first calculate the energy
dependent LDOS images due to a single extended defect and show how
the pattern of the image changes as the energy varies. Then we
demonstrate that the experimentally observed checkerboard pattern
for the LDOS could be understood as a result of interferences
among randomly distributed defects. Our results clearly indicate
the dominant LDOS modulation along $45^0$ to the Cu-O bonds at lower
energy would transform to that oriented parallel to the Cu-O bonds
at higher energy. This is also consistent with the STM
experiments [1,2].

The Hamiltonian describing the scattering of quasiparticles from
$M$ impurities with local modifications of both hopping and
pairing parameters in a d-wave superconductor can be written as
$$H=H_{\rm BCS}+H_{\rm imp},\eqno{(1)}$$
where
$$H_{\rm BCS}=\sum_{{\bf k} \sigma}(\epsilon_{\bf
k}-\mu)c^\dag_{{\bf k}\sigma}c_{{\bf k}\sigma}+\sum_{\bf
k}\Delta_{\bf k} (c^\dag_{{\bf k}\uparrow}c^\dag_{-{\bf
k}\downarrow} +c_{-{\bf k}\downarrow}c_{{\bf k}\uparrow}),$$
$$H_{\rm imp}=\sum_{<i,j>, \sigma}\delta
t_{ij}c^\dag_{i\sigma}c_{j\sigma} +\sum_{<i,j>}\delta \Delta_{ij}
(c^\dag_{i\uparrow}c^\dag_{{j}\downarrow}
+c_{j\downarrow}c_{i\uparrow})$$
$$+\sum_{i=1}^{M}[(V_{si}+V_{mi})c^\dag_{{\bf R}_i\uparrow}c_{{\bf
R}_i\uparrow} +(V_{si}-V_{mi})c^\dag_{{\bf R}_i\downarrow}c_{{\bf
R}_i\downarrow}].$$
Here $\mu$ is the chemical potential to be determined by doping,
$\epsilon_k=t_1({\rm cos}k_x+{\rm
cos}k_y)/2+t_2{\rm cos}k_x{\rm cos}k_y+t_3({\rm cos}2k_x+{\rm
cos}2k_y)/2 +t_4({\rm cos}2k_x{\rm cos}k_y+{\rm cos}k_x{\rm
cos}2k_y)/2+t_5{\rm cos}2k_x{\rm cos}2k_y$, where
$t_{1-5}=-0.5951, 0.1636, -0.0519, -0.1117, 0.0510$ (eV). The band
parameters are taken from those of Norman {\it et al.} [7] for
Bi$_2$Sr$_2$CaCu$_2$O$_{8+\delta}$, and the lattice constant $a$
is set as $a=1$. The order parameter away from the impurity is
given by $\Delta_{\bf k}=\Delta_0({\rm cos}k_x-{\rm cos}k_y)/2$.

Without loss of generality, at the impurity or defect site $R_i$,
we assume an on-site potential consisting of a nonmagnetic part,
$V_{si}$, and a magnetic part, $V_{mi}$. The defect also induces a
weak local modification in the hopping, $\delta t_{i}$, to the
nearest neighbor sites, and a suppression of the superconductivity
order parameter on the four bonds connected to the impurity site,
$\delta \Delta_{1i}$, and on the other twelve bonds connected to
the nearest neighbor sites, $\delta\Delta_{2i}$.

The Hamiltonian (1) with a single impurity has in fact been
successfully applied by authors in Ref. 8 to explain the resonant
STM spectra for Ni impurities in
Bi$_2$Sr$_2$CaCu$_2$O$_{8+\delta}$. However, in the present case,
no resonances in LDOS have been observed in the recent STM
experiments [1, 2]. So it is reasonable to assume that the on-site
potentials ($V_{si}$ and $V_{mi}$) and the modifications in
hopping and pairing parameters ($\delta t_{i}$, $\delta
\Delta_{1i}$ and $\delta \Delta_{2i}$) are all weak and have
approximately the same order of magnitude. This model has been
applied in Ref. 4 to explain the experimental observed Fourier
transform of the LDOS at different energies and dopings in the
first Brillouin zone [1, 2].

The Hamiltonian (1) can be solved by the standard Bogoliubov
transformation plus Green's function technique. When $\delta
t_{i}, \delta\Delta_{1i}, \delta\Delta_{2i}, V_{si}$ and $V_{mi}$
are all small, keeping the leading term in the T-matrix approach
should be good approximation. In such an approximation, the
resonant states due to the impurity are eliminated and the LDOS
change in real space due to these defects can be shown to have the
following form
$$\delta \rho({\bf r},\omega)=-\frac{2}{\pi
N^2}\sum_{i=1}^{M} \sum_{{\bf k},{\bf k}^\prime}
\sum_{\nu,\nu^\prime=0,1}cos[({\bf k}-{\bf k}^\prime) \cdot ({\bf
r}-{\bf R}_i)] $$
$$ \times \{[2\delta t_{i} A({\bf k}, {\bf
k}^\prime)+V_{si}] \alpha_{\nu\nu^\prime}({\bf k}, {\bf
k}^\prime)$$
$$ +2[\delta\Delta_{1i} B({\bf k}, {\bf k}^\prime)+
\delta\Delta_{2i} C({\bf k}, {\bf k}^\prime)]
\beta_{\nu\nu^\prime}({\bf k}, {\bf k}^\prime)\}$$
$$ \times {\rm
Im}[ G^0_{{\bf k}\nu}(i\omega_n)G^0_{{\bf k}^\prime\nu^\prime}
(i\omega_n)]|_{i\omega_n\rightarrow \omega+i0^+}, \eqno{(2)}$$
where $N$ is the number of sites in the lattice, $A({\bf k}, {\bf
k}^\prime) =cosk_x+cosk_y+cosk_x^\prime+cosk_y^\prime$, $B({\bf
k}, {\bf k}^\prime) =cosk_x-cosk_y+cosk_x^\prime-cosk_y^\prime$,
$C({\bf k}, {\bf k}^\prime)
=cos(k_x-2k_x^\prime)-cos(k_x-k_x^\prime-k_y^\prime)
-cos(k_x-k_x^\prime+k_y^\prime)+cos(k_y-k_x^\prime-k_y^\prime)
+cos(k_y-k_y^\prime+k_x^\prime)-cos(k_y-2k_y^\prime)
+cos(2k_x-k_x^\prime)+cos(k_x+k_y-k_y^\prime)-
cos(k_x+k_y-k_x^\prime)-cos(2k_y-k_y^\prime)
+cos(k_x-k_y+k_y^\prime)-cos(k_x-k_y-k_x^\prime)$,
$\alpha_{\nu\nu^\prime}({\bf k}, {\bf k}^\prime)=\xi^2_{{\bf
k}\nu} \xi^2_{{\bf k}^\prime\nu^\prime}-(-1)^{\nu+\nu^\prime}
\xi_{{\bf k}\nu}\xi_{{\bf k}\nu+1}\xi_{{\bf k}^\prime\nu^\prime}
\xi_{{\bf k}^\prime\nu^\prime+1}$, $\beta_{\nu\nu^\prime}({\bf k},
{\bf k}^\prime)=(-1)^\nu\xi_{{\bf k}\nu}\xi_{{\bf k}\nu+1}
\xi^2_{{\bf k}^\prime\nu^\prime}+(-1)^{\nu^\prime}\xi^2_{{\bf
k}\nu} \xi_{{\bf k}^\prime\nu^\prime}\xi_{{\bf
k}^\prime\nu^\prime+1}$, $G^0_{{\bf
k}\nu}(i\omega_n)=1/[i\omega_n-(-1)^\nu E_{{\bf k}}]$ is the bare
Green's function, $E_{{\bf k}}=\sqrt{(\epsilon_{\bf k}-\mu)^2
+\Delta^2_{{\bf k}}}$, $\xi^2_{{\bf k}\nu}=[1+(-1)^\nu
(\epsilon_{\bf k}-\mu)/E_{{\bf k}}]/2$, and $\xi_{{\bf k}0}
\xi_{{\bf k}1}=\Delta_{{\bf k}}/(2E_{{\bf k}})$.

We note that $V_{mi}$ is absent from Eq. (2) because there is no
first order contribution from the magnetic potential. Obviously,
the total LDOS change $\delta \rho({\bf r},\omega)$ is a summation
of those due to individual impurity. In the present study, we base
our numerical calculation on a finite lattice of $800\times 800$
sites. For simplicity, we choose $2\delta
t_i=V_{si}=-2\delta\Delta_{1i}=-4\delta\Delta_{2i}$, and assume
that all these parameters are small such that the first order
T-matrix approximation is valid. In our calculation, we also take
the chemical potential $\mu=-0.1238$ corresponding to the optical
doping ($15\%$) and introduce a finite lifetime broadening
$\gamma=2$ meV to the quasiparticle Green's function to smooth our
data points by replacing $\omega +i0^+$ with $\omega +i\gamma$ in
Eq. (2).

According to Eq. (2), We plot the images of the LDOS change due to
a single defect ($M=1$) located at the center of a $20\times 20$
square lattice for different energies in Fig. 1. It is easy to see
from the images at $\omega=0$ and -12meV that the LDOS
modulations orient parallel to $(\pm 1, \pm 1)$ directions ($45^0$
to Cu-O bonds). When the energy becomes more negative at $\omega
=-16, -20$ and $-25$meV, the LDOS modulation clearly changes to $(\pm
1, 0)$ or $(0, \pm 1)$ (along the Cu-O bonds) directions. At
$\omega$ =12 and 16meV, the LDOS modulations are strongly aligned
along the directions of $45^0$ to Cu-O bonds. For higher energies
at $\omega =20$ and 25meV, the modulations along
the Cu-O bonds begin to show up, and they may become dominant at
higher energies. It is apparent that the LDOS images are
asymmetric with respect to $\omega $ =0. We also note that
with increasing energy $|\omega|$, the region of the LDOS
modulation along $(\pm 1, \pm 1)$ directions becomes smaller
while that along $(\pm 1, 0)$ or $(0, \pm 1)$ directions
becomes larger.

In order to understand the LDOS modulations near a single
impurity, we present the LDOS variations along (1, 0) and (1, 1)
directions in Fig. 2. We can see that the LDOS on the impurity site
has the maximum values at energy $\omega <$ 12meV. At and above 12meV,
the LDOS  near the nearest neighboring site
to the impurity have the maximum values. Far away the
impurity(about 20a), the LDOS modulations vanish.
For $\omega <-12 $ meV, the LDOS along
(1,0) direction show strong and long distance oscillations
while those along (1,1) direction show rather week and short
distance oscillations. This is the reason why
the LDOS images shown in Fig.1 in this energy region have
modulation vectors clearly along Cu-O bonds. At $\omega$= -12, 0
and 12meV, both modulations with approximately equal weight are
present in the LDOS. As a result, the LDOS images with modulations
along the directions of $45^0$ to Cu-O bonds can also be seen in
this region. For $\omega >$12meV, the strength of the modulation
along (1,0) direction begins to outgrow that along (1,1)
direction. We expect that dominant modulation should be along
(1,0) direction, as the energy gets even higher. In fact there are
total six nonequivalent charge modulation vectors [2,4], not all
of them can be clearly identified in our real space studies.

We have discussed the LDOS modulations due to a single impurity.
In fact, there should be randomly distributed defects in the
experimental sample. In order to understand the effect of the
disorder, we choose the defect concentration to be $1\%$ and the
defects are described by $H_{\rm imp}$ in Eq.(1). In our
simulation, we first produce a random distribution for the defects
and then calculate the LDOS changes in a $20\times 20$ square lattice
similar to that of the experiments [1, 2].

Graphs in Fig. 3 show the LDOS images at 9 different energies for
such a distribution of defects. As a consequence of quasiparticle
interference by these defects, checkerboard patterns in the LDOS show
up in our numerical simulations. It is evident from Fig. 3 that
the LDOS modulation or the orientation of the checkerboard pattern is
parallel to the $45^0$ direction from the Cu-O bonds at small
$|\omega|$, while at larger $|\omega|$, the pattern tends to
orient along the direction of the Cu-O bonds. The changing of the
orientation is particularly apparent when $\omega$ becomes more
negative. This conclusion is consistent with the STM experiments
[1, 2].

In summary, we have studied the LDOS change induced by disordered
and weak defects using Bogoliubov transformation and the the
Green's function technique. The obtained LDOS images due to
randomly distributed defects exhibit checkerboard patterns which
are similar to those observed in the STM experiments [1, 2]. With
increasing energy $|\omega|$, the LDOS modulation along $(\pm 1,
\pm 1)$ direction tends to orient itself in the $(\pm 1, 0)$ or
$(0, \pm 1)$ direction. This modulation transformation has also
been seen in these experiments. Combining our previous work [4],
we conclude that the STM images [1, 2] can be qualitatively
understood in the present theory. For the charge density wave
order observed in Ref. [9], it is believed to be the dimerization
hopping and transverse pairing modulations [10, 11, 12].
However, the origin of such a static order needs to be further
studied.

We wish to thank Professor S. H. Pan for useful discussion, and
Professor J. C. Davis for sending us Ref. [2] before its
publication. This work has been supported by the Texas Center for
Superconductivity at the University of Houston and by the Robert
A. Welch Foundation.

\begin{figure}

\epsfxsize=1cm \vspace{2cm} \narrowtext \caption {The LDOS change
$\delta \rho({\bf r},\omega)$ at different
energy due to a single impurity at the center
of a $20\times 20$ square.}
\end{figure}

\begin{figure}

\caption {The LDOS change $\delta \rho({\bf r},\omega)$ versus the
distance $|{\bf r}|$ to the single impurity along $(1, 0)$ and
$(1, 1)$ directions at different energy.}
\end{figure}

\begin{figure}
\caption {The LDOS change $\delta \rho({\bf r},\omega)$
at different energy due to the
random impurities in a $20\times 20$ square.}
\end{figure}

\end{multicols}

\end{document}